# Performance Analysis of Connection Admission Control Scheme in IEEE 802.16 OFDMA Networks

Abdelali EL BOUCHTI, Said EL KAFHALI and Abdelkrim HAQIQ

Computer, Networks, Mobility and Modeling laboratory
e- NGN research group, Africa and Middle East
FST, Hassan 1st University, Settat, Morocco
Emails: {a.elbouchti, kafhalisaid, ahaqiq} @gmail.com

*Abstract*—**IEEE 802.16 OFDMA (Orthogonal Frequency Division Multiple Access) technology has emerged as a promising technology for broadband access in a Wireless Metropolitan Area Network (WMAN) environment. In this paper, we address the problem of queueing theoretic performance modeling and analysis of OFDMA under broad-band wireless networks. We consider a single-cell IEEE 802.16 environment in which the base station allocates subchannels to the subscriber stations in its coverage area. The subchannels allocated to a subscriber station are shared by multiple connections at that subscriber station. To ensure the Quality of Service (QoS) performances, a Connection Admission Control (CAC) scheme is considered at a subscriber station. A queueing analytical framework for these admission control schemes is presented considering OFDMA-based transmission at the physical layer. Then, based on the queueing model, both the connection-level and the packet-level performances are studied and compared with their analogues in the case without CAC. The connection arrival is modeled by a Poisson process and the packet arrival for a connection by a two-state Markov Modulated Poisson Process (MMPP). We determine analytically and numerically different performance parameters, such as connection blocking probability, average number of ongoing connections, average queue length, packet dropping probability, queue throughput and average packet delay.**

## I. INTRODUCTION

The evolution of the IEEE 802.16 standard [14] has spurred tremendous interest from the network operators seeking to deploy high performance, cost-effective broadband wireless networks. With the aid of the Worldwide Interoperability for Microwave Access (WiMAX) organization [1], several commercial implementations of WiMAX cellular networks have been launched, based on OFDMA for non-line-of-sight applications. The IEEE 802.16/WiMAX [2] can offer a high data rate, low latency, advanced security, quality of service (QoS), and low-cost deployment.

OFDMA is a promising wireless access technology for the next generation broad-band packet networks. With OFDMA, which is based on orthogonal frequency division multiplexing (OFDM), the wireless access performance can be substantially improved by transmitting data via multiple parallel channels, and also it is robust to inter-symbol interference and frequency-selective fading. OFDMA has been adopted as the physical layer transmission technology for IEEE 802.16/WiMAX-based broadband wireless networks. Although the IEEE 802.16/WiMAX standard [12] defines the physical layer specifications and the Medium Access Control (MAC) signaling mechanisms, the radio resource management methods such as those for Connection Admission Control (CAC) and dynamic bandwidth adaptation are left open. However, to guarantee QoS performances (e.g., call blocking rate, packet loss, and delay), efficient admission control is necessary in a WiMAX network at both the subscriber and the base stations.

The admission control problem was studied extensively for wired networks (e.g., for ATM networks) and also for traditional cellular wireless systems. The classical approach for CAC in a mobile wireless network is to use the guard channel scheme [5] in which a portion of wireless resources (e.g., channel bandwidth) is reserved for handoff traffic. A more general CAC scheme, namely, the fractional guard scheme, was proposed [13] in which a handoff call/connection is accepted with a certain probability. To analyze various connection admission control algorithms, analytical models based on continuous-time Markov chain, were proposed [4]. However, most of these models dealt only with call/connection-level performances (e.g., new call blocking and handoff call dropping probabilities) for the traditional voice-oriented cellular networks. In addition to the connection-level performances, packet-level (i.e., in-connection) performances also need to be considered for data-oriented packet-switched wireless networks such as WiMAX networks.

An earlier relevant work was reported by the authors in [10]. They considered a similar model in OFDMA based-IEEE 802.16 but they modeled both the connection-level and packet-level by tow different Poisson processes and they compared various QoS measures of CAC schemes. In [15], the authors proposed a Discrete-Time Markov Chain (DTMC) framework based on a Markov Modulated Poisson Process (MMPP) traffic model to analyze VoIP performance. The MMPP processes are very suitable for formulating the multi-user VoIP traffic and capturing the interframe dependency between consecutive frames.

In this paper, we present a connection admission control scheme for a multi-channel and multi-user OFDMA network, in which the concept of guard channel is used to limit the number of admitted connections to a certain threshold. A queueing analytical model is developed based on a three-DTMC which captures the system dynamics in terms of the number of connections and queue status. We assume that the connection arrival and the packet arrival for a connection follow a Poisson process and a two-state MMPP process respectively. Based on this model, various performance parameters such as connection blocking probability, average number of ongoing connections, average queue length, probability of packet dropping due to lack of buffer space, queue throughput, and average queueing delay are obtained. The numerical results reveal the comparative performance characteristics of the CAC and the without CAC algorithms in an OFDMA-based WiMAX network.

The remainder of this paper is organized as follows. Section II describes the system model including the objective of CAC policy. The formulation of the analytical model for connection admission control is presented in Section III. In section IV we determine analytically different performance parameters. Numerical results are stated in Section V. Finally, section VI concludes the paper.

## II. MODEL DESCRIPTION

### A. System model

We consider a single cell in a WiMAX network with a base station and multiple subscriber stations (Figure 1). Each subscriber station serves multiple connections. Admission control is used at each subscriber station to limit the number of ongoing connections through that subscriber station. At each subscriber station, traffic from all uplink connections are aggregated into a single queue [11]. The size of this queue is finite (i.e., *L* packets) in which some packets will be dropped if the queue is full upon their arrivals. The OFDMA transmitter at the subscriber station retrieves the head of line packet(s) and transmits them to the base station. The base station may allocate different number of subchannels to different subscriber stations. For example, a subscriber station with higher priority could be allocated more number of subchannels.

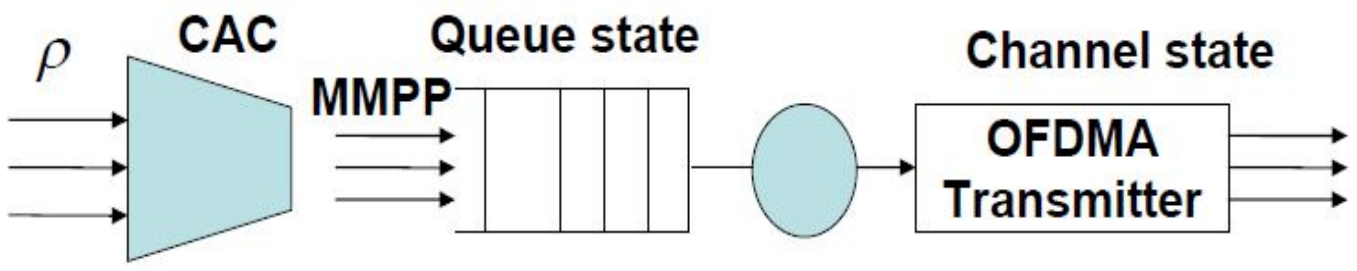

Figure 1. System model

### B. CAC Plicy

The main objective of a CAC mechanism is to limit the number of ongoing connections/flows so that the QoS performances can be guaranteed for all the ongoing connections. Then, the admission control decision is made to accept or reject an incoming connection. To ensure the QoS performances of the ongoing connections, the following CAC scheme for subscriber stations are proposed. A threshold C is used to limit the number of ongoing connections. When a new connection arrives, the CAC module checks whether the total number of connections including the incoming one is less than or equal to the threshold C. If it is true, then the new connection is accepted, otherwise it is rejected.

## III. FORMULATION OF THE ANALYTICAL MODEL

### A. Formulation of the Queueing Model

An analytical model based on DTMC is presented to analyze the system performances at both the connection-level and at the packet-level for the connection admission schemes described before. We assume that packet arrival for a connection follows a two-state MMPP process [3] which is identical for all connections in the same queue. The connection inter-arrival time and the duration of a connection are assumed to be exponentially distributed with average $1/\rho$ and $1/\mu$, respectively.

An MMPP is a stochastic process in which the intensity of a Poisson process is defined by the states of a Markov chain. That is, the Poisson process can be modulated by a Markov chain. As mentioned before, an MMPP process can be used to model time-varying arrival rates and can capture the inter-frame dependency between consecutive frames ([6], [7], [8]). The transition rate matrix and the Poisson arrival rate matrix of the two-state MMPP process can be expressed as follows:

$$Q_{MMPP} = \begin{pmatrix} -q_{01} & q_{01} \\ q_{10} & -q_{10} \end{pmatrix}, \ \Lambda = \begin{pmatrix} \lambda_0 & 0 \\ 0 & \lambda_1 \end{pmatrix} \qquad (1)$$

The steady-state probabilities of the underlying Markov chain are given by:

$$(\pi_{MMPP,0}, \pi_{MMPP,1}) = \left(\frac{q_{10}}{q_{01}+q_{10}}, \frac{q_{01}}{q_{01}+q_{10}}\right) \qquad (2)$$

The mean steady state arrival rate generated by the MMPP is:

$$\lambda_{MMPP} = \pi_{MMPP}\lambda^T = \frac{q_{10}\lambda_0 + q_{01}\lambda_1}{q_{01}+q_{10}} \qquad (3)$$

where $\lambda^T$ is the transpose of the row vector $\lambda = (\lambda_0, \lambda_1)$.

The state of the system is described by the process $X_t = (X, X_t^1, X_t^2)$, where $X$ is the state of an irreducible continuous time Markov chain and $X_t^1$ (respectively $X_t^2$) is the number of packets in the aggregated queue (the number of ongoing connections) at the end of every time slot *t*.

Thus, the state space of the system is given by:

$$E = \{(i, j, k) / i \in \{0,1\},\ 0 \le j \le L,\ k \ge 0\}\ .$$

For the CAC algorithm, the number of packet arrivals depends on the number of connections. The state transition diagram is shown in (Figure 2). Here, $(\lambda_0, \lambda_1)$ and $\rho$ denote rates and not probabilities.

Note that the probability that *n* Poisson events with average rate $\rho$ occur during an interval T can be obtained as follows:

$$f_n(\rho) = \frac{e^{-\rho T}(\rho T)^n}{n!} \tag{4}$$

This function is required to determine the probability of both connection and packet arrivals.

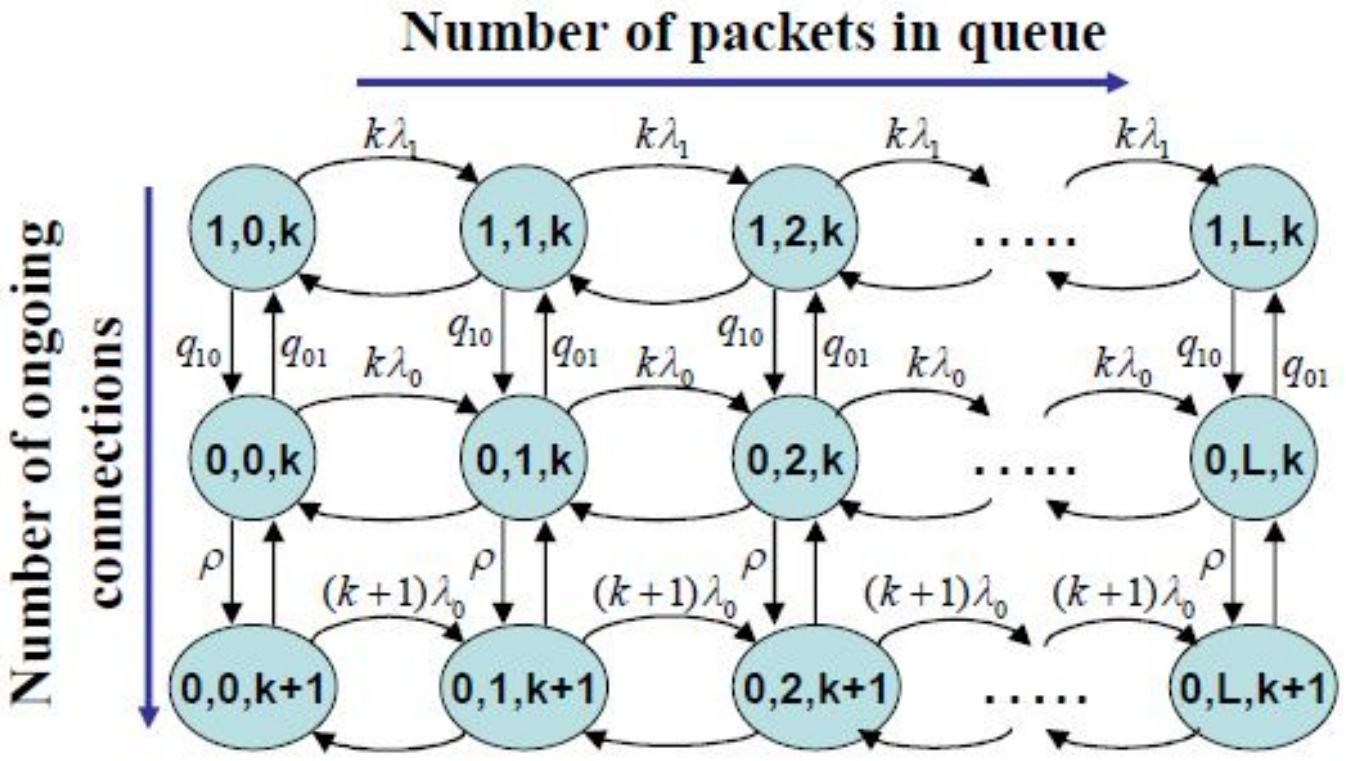

Figure 2. State transition diagram of discrete time Markov chain.

## B. CAC Algorithm

In this case, the transition matrix $Q$ for the number of connections in the system can be expressed as follows:

$$Q = \begin{bmatrix} q_{0,0} & q_{0,1} & & & \\ q_{0,1} & q_{1,1} & q_{1,2} & & \\ \ddots & \ddots & \ddots & & \\ & & q_{C-2,C-1} & q_{C-1,C-1} & q_{C-1,C} \\ & & & q_{C-1,C} & q_{C,C} \end{bmatrix} \tag{5}$$

where each row indicates the number of ongoing connections. As the length of a frame *T* is very small compared with connection arrival and departure rates, we assume that the maximum number of arriving and departing connections in a frame is one. Therefore, the elements of this matrix can be obtained as follows:

$$\begin{aligned} q_{k,k+1} &= f_1(\rho)\times(1-f_1(k\mu)), \ \ \text{k=0,1,...,C-1} \\ q_{k,k-1} &= (1-f_1(\rho))\times f_1(k\mu), \ \ \text{k=1,2,...,C} \\ q_{k,k} &= f_1(\rho)\times f_1(k\mu)+(1-f_1(\rho))\times(1-f_1(k\mu)), \ \ \text{k=0,1,...,C} \end{aligned} \tag{6}$$

where $q_{k,k+1}$, $q_{k,k-1}$ and $q_{k,k}$ represent the cases that the number of ongoing connections increases by one, decreases by one, and does not change, respectively.

## C. Transition Matrix for the Queue

The transition matrix *P* of the entire system can be expressed as follows. The rows of matrix *P* represent the number of packets (*j*) in the queue.

$$P = \begin{bmatrix} p_{0,0} & \cdots & p_{0,A} & & & & \\ \vdots & \vdots & \ddots & & \ddots & & \\ p_{R,0} & \cdots & p_{R,R} & \cdots & p_{R,R+A} & & \\ \ddots & \ddots & \ddots & \ddots & \ddots & & \ddots \\ & p_{j,j-R} & \cdots & p_{j,j} & \cdots & & p_{j,j+R} \\ & \ddots & \ddots & \ddots & \ddots & & \ddots \end{bmatrix} \tag{7}$$

Matrices $p_{j,j'}$ represent the changes in the number of packets in the queue (i.e., the number of packets in the queue changing from *j* in the current frame to $j'$ in the next frame). We first establish matrices $\text{v}_{(i,j),(i,j')}$, where the diagonal elements of these matrices are given as follows. For $r \in \{0,1,2,...,D\}$ and $n \in \{0,1,2,...,(k\times A)\}$, $l = 1,2,...,D$, and $m = 1,2,...,(k\times A)$. The non-diagonal elements of $\text{v}_{(i,j),(i,j')}$ are all zero.

$$\begin{aligned} \left[\text{v}_{(i,j);(i,j-l)}\right]_{k+1,k+1} &= \sum_{n-r=l} f_n(k\lambda_i)[R]_r \\ \left[\text{v}_{(i,j);(i,j+m)}\right]_{k+1,k+1} &= \sum_{r-n=m} f_n(k\lambda_i)[R]_r \\ \left[\text{v}_{(i,j);(i,j)}\right]_{k+1,k+1} &= \sum_{r=n} f_n(k\lambda_i)[R]_r \end{aligned} \tag{8}$$

Here *A* is the maximum number of packets that can arrive from one connection in one frame, *R* indicates the maximum number of packets that can be transmitted in one frame and *D* is the maximum number of packets that can be transmitted in one frame by all of the allocated subchannels allocated to that particular queue and it can be obtained from $D = \min(R, j)$. This is due to the fact that the maximum number of transmitted packets depends on the number of packets in the queue and the maximum possible number of transmissions in one frame. Note that, $\left[\text{v}_{(i,j);(i,j-l)}\right]_{k+1,k+1}$, $\left[\text{v}_{(i,j);(i,j+m)}\right]_{k+1,k+1}$ and $\left[\text{v}_{(i,j);(i,j)}\right]_{k+1,k+1}$ represent the probability that the number of packets in the queue increases by *n*, decreases by *m*, and does not change, respectively, when there are *k* ongoing connections. Here, $[\text{v}]_{i,j}$ denotes the element at row *i* and column *j* of matrix v, and these elements are obtained based on the assumption that the packet arrivals for the ongoing connections are independent of each other.

Finally, we obtain the matrices $p_{j,j'}$ by combining both the connection-level and the queue-level transitions as follows:

$$p_{j,j'} = Q\text{v}_{(i,j),(i,j')} \tag{9}$$

## IV. Performance Parameters

In this section, we determine the connection-level and the packet-level performance parameters (i.e., connection blocking probability, average number of ongoing connections in the system, and average queue length) for the CAC scheme.

These performance parameters can be derived from the steady state probability vector of the system states $\pi$ , which is obtained by solving $\pi P = \pi$ and $\pi 1 = 1$, where $1$ is a column matrix of ones.

Also, the size of the matrix *P* needs to be truncated at *L* (i.e., the maximum number of packets in the queue) for the scheme.

The steady-state probability, denoted by $\pi(i, j, k)$ for the state that there are *k* connections and $j \in \{0,1,...,L\}$ packets in the queue, can be extracted from matrix $\pi$ as follows:

$$\pi(i, j, k) = [\pi]_{i \times j \times ((C+1)+k)}, \quad i = 0,1;\ k = 0,1,...,C \tag{10}$$

### A. Connection Blocking Probability

This performance parameter indicates that an arriving connection will be blocked due to the admission control decision. It indicates the accessibility of the wireless service and can be obtained as follows:

$$p_{block} = \sum_{i=0}^{1} \sum_{j=0}^{L} \pi(i, j, C). \tag{11}$$

The above probability refers to the probability that the system serves the maximum allowable number of ongoing connections.

### B. Average Number of Ongoing Connections

It can be obtained as

$$N_k = \sum_{i=0}^{1} \sum_{j=0}^{L} \sum_{k=0}^{C} k.\pi(i, j, k) \tag{12}$$

### C. Average Queue Length Average

It is given by

$$N_j = \sum_{i=0}^{1} \sum_{k=0}^{C} \sum_{j=0}^{L} j.\pi(i, j, k) \tag{13}$$

### D. Packet Dropping Probability

It refers to the probability that an incoming packet will be dropped due to the unavailability of buffer space. It can be derived from the average number of dropped packets per frame. Given that there are *j* packets in the queue and the number of packets in the queue increases by v, the number of dropped packets is $m-(L-j)$ for $m > L-j$ , and zero otherwise. The average number of dropped packets per frame is obtained as follows:

$$N_{drop} = \sum_{i=0}^{1} \sum_{k=1}^{C} \sum_{j=0}^{L} \sum_{m=L-j+1}^{A} \left( \sum_{l=1}^{C} [p_{j,j+m}]_{k,l} \right).(m-(L-j)).\pi(i, j, k) \tag{14}$$

where the term $\left( \sum_{l=1}^{C} [p_{j,j+m}]_{k,l} \right)$ indicates the total probability that the number of packets in the queue increases by $m$ at every arrival phase. Note that, we consider probability $p_{j,j+m}$ rather than the probability of packet arrival as we have to consider the packet transmission in the same frame as well.

After calculating the average number of dropped packets per frame, we can obtain the probability that an incoming packet is dropped as follows:

$$p_{drop} = \frac{N_{drop}}{\overline{\lambda}} \tag{15}$$

where $\overline{\lambda}$ is the average number of packet arrivals per frame and it can be obtained from

$$\overline{\lambda} = \lambda_{MMPP} N_k . \tag{16}$$

### E. Queue throughput

It measures the number of packets transmitted in one frame and can be obtained from

$$\eta = \lambda_{MMPP} (1 - p_{drop}). \tag{17}$$

### F. Average Packet Delay

It is defined as the number of frames that a packet waits in the queue since its arrival before it is transmitted. We use Little's law [9] to obtain average delay as follows:

$$D = \frac{N_j}{\eta} \tag{18}$$

## V. Numerical Results

In this section we present the numerical results of CAC scheme. We use the Matlab software to solve numerically and to evaluate the various performance parameters.

### A. Parameter Setting

As in [10], we consider one queue (which corresponds to a particular subscriber station) for which five subchannels are allocated and we assume that the average SNR is the same for all of these subchannels. Each subchannel has a bandwidth of 160 kHz. The length of a subframe for downlink transmission is one millisecond, and therefore, the transmission rate in one subchannel with rate ID = 0 (i.e., BPSK modulation and coding rate is 1/2) is 80 kbps. We assume that the maximum number of packets arriving in one frame for a connection is limited to 30 (i.e., A = 30).

For our scheme, the value of the threshold C is varied according to the evaluation scenarios.

For performance comparison, we also evaluate the queueing performance in the absence of CAC mechanism. For the case without CAC, we truncate the maximum number of ongoing connections at 25 (i.e. $C_{tr} = 25$) so that $\pi(i,j,C_{tr}) < 2.10^{-4}$, $\forall i,j$. The average duration of a connection is set to ten minutes (i.e., $\mu$ = 10) for all the evaluation scenarios. The queue size is 150 packets (i.e., $L$ = 150). The parameters are set as follows: The connection arrival rate is 0.4 connections per minute. Packet arrival rate per connection is one packet per frame for state 0 of MMPP process and two packets per frame for state 1 of MMPP process. Average SNR on each subchannel is 5 dB.
Note that, we vary some of these parameters depending on the evaluation scenarios whereas the others remain fixed.

### *B. Performance of CAC policy*

We first examine the impact of connection arrival rate on connection-level performances. Variations in average number of ongoing connections and connection blocking probability with connection arrival rate are shown in Figures 3 and 4, respectively. As expected, when the connection arrival rate increases, the number of ongoing connections and connection blocking probability increase.

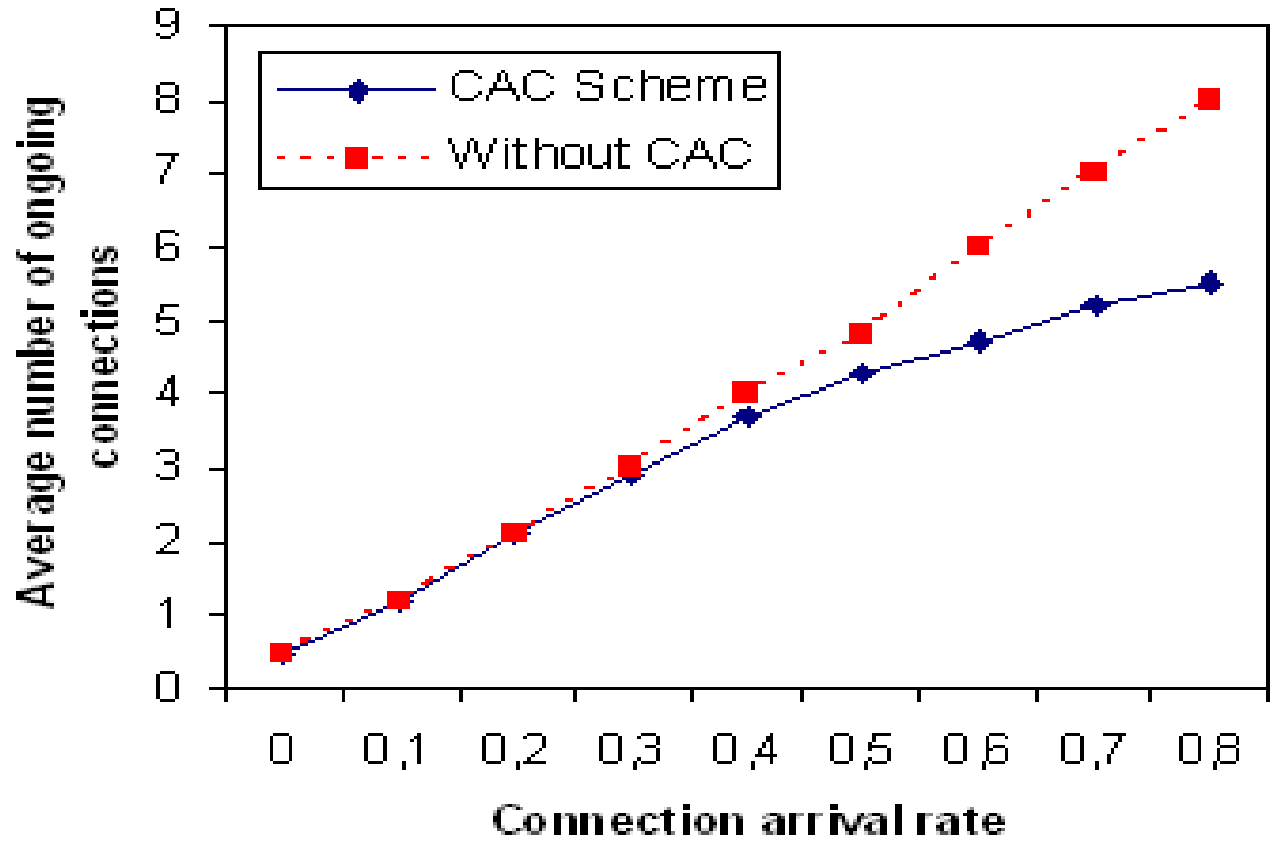

Figure 3: Average number of ongoing connections under different connection arrival rates.

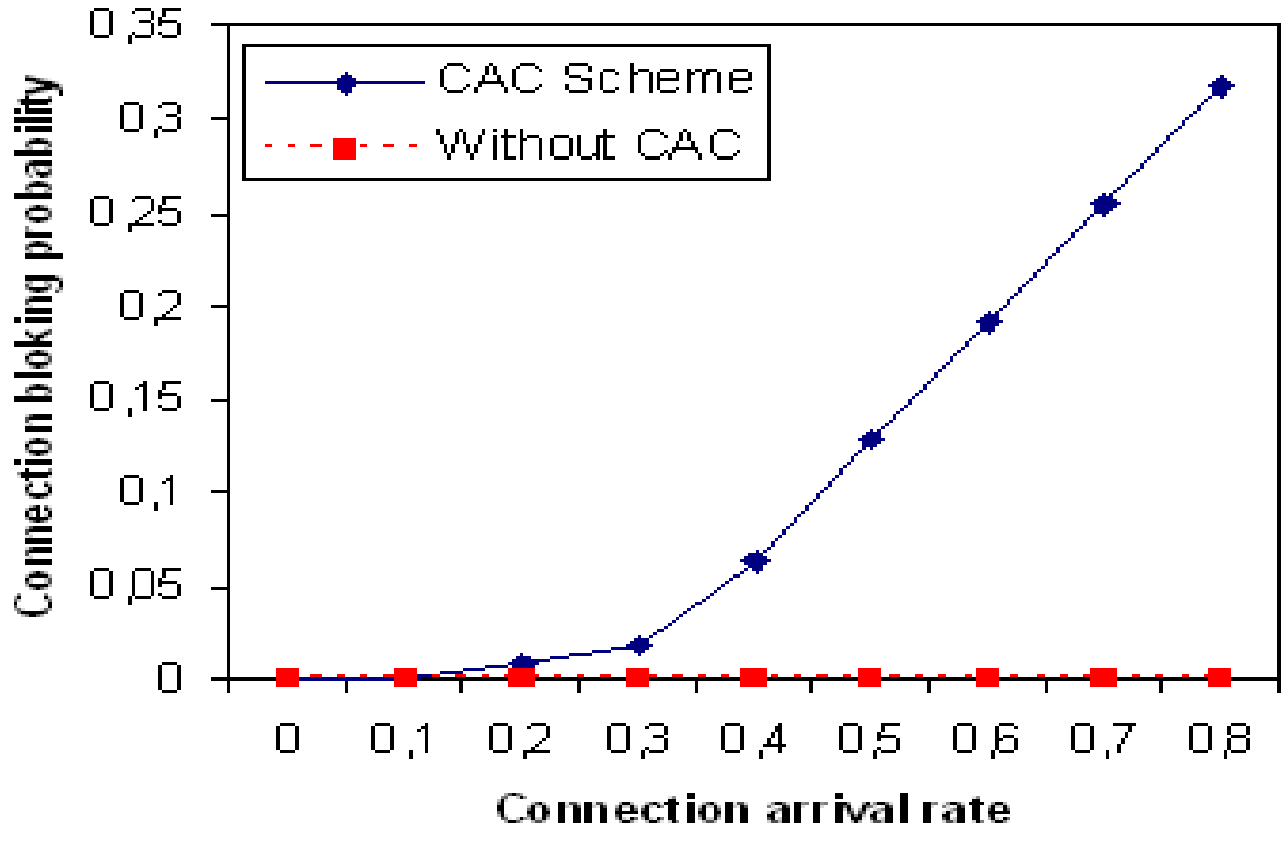

Figure 4: Connection blocking under different connection arrival rates.

The packet-level performances under different connection arrival rates are shown in Figures 5 through 8 for average number of packets in the queue, packet dropping probability, queue throughput, and average queueing delay, respectively. These performance parameters are significantly impacted by the connection arrival rate. Because the CAC scheme limits the number of ongoing connections, packet-level performances can be maintained at the target level. In this case, the CAC scheme results in better packet-level performances compared with those without CAC scheme.

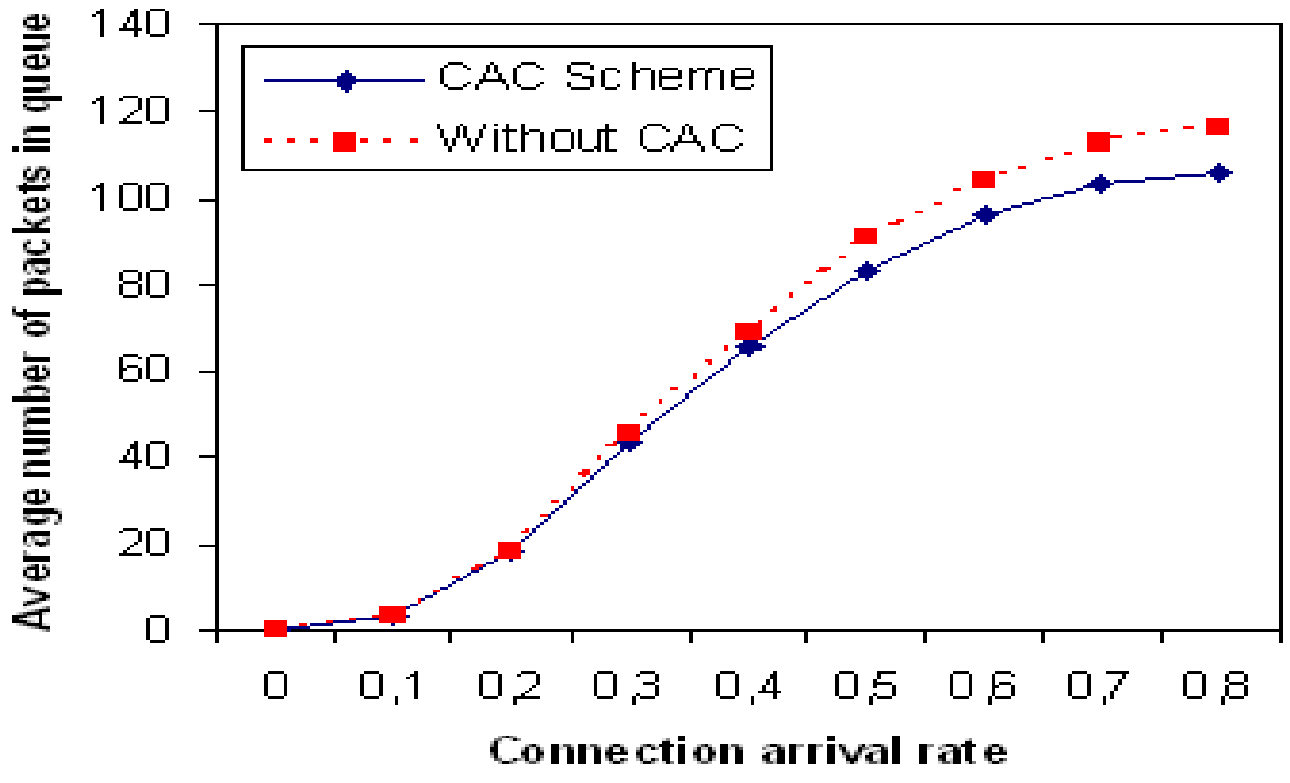

Figure 5: Average number of packets in queue under different connection rates.

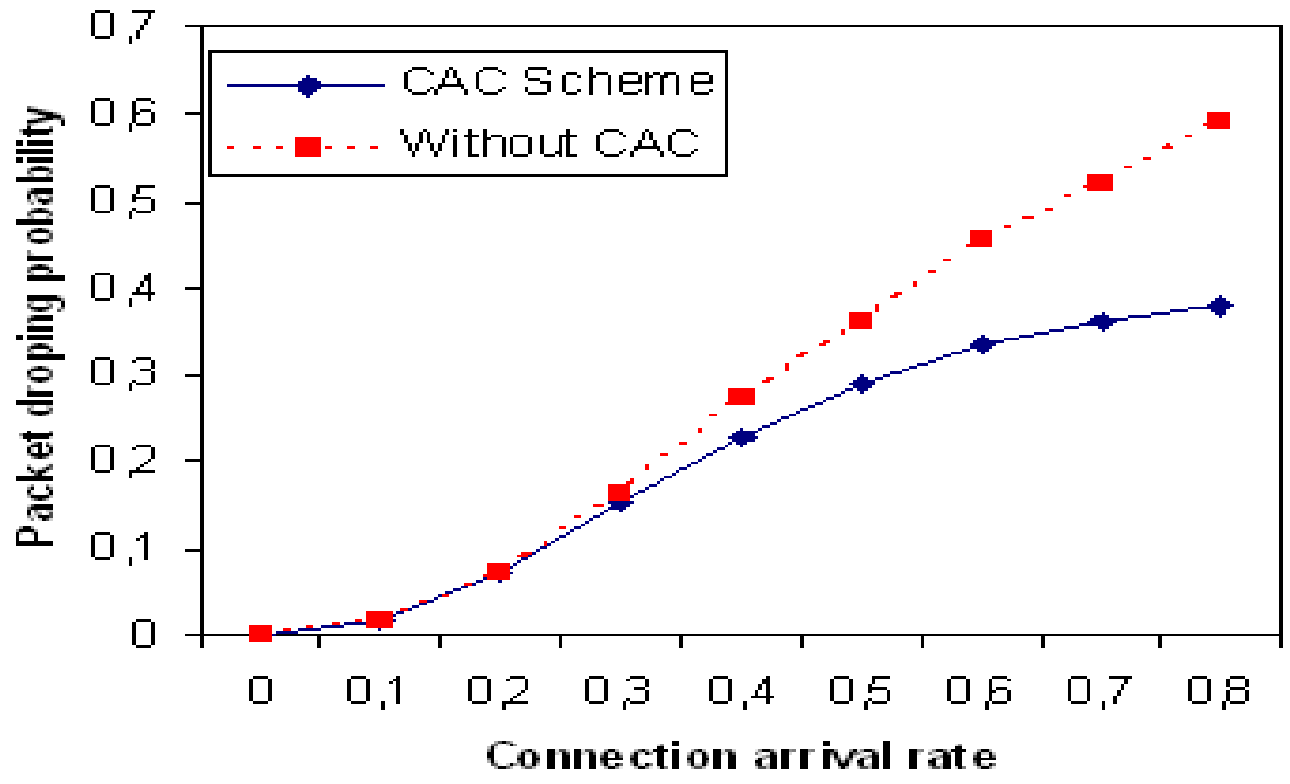

Figure 6: Packet dropping under different connection arrival rates.

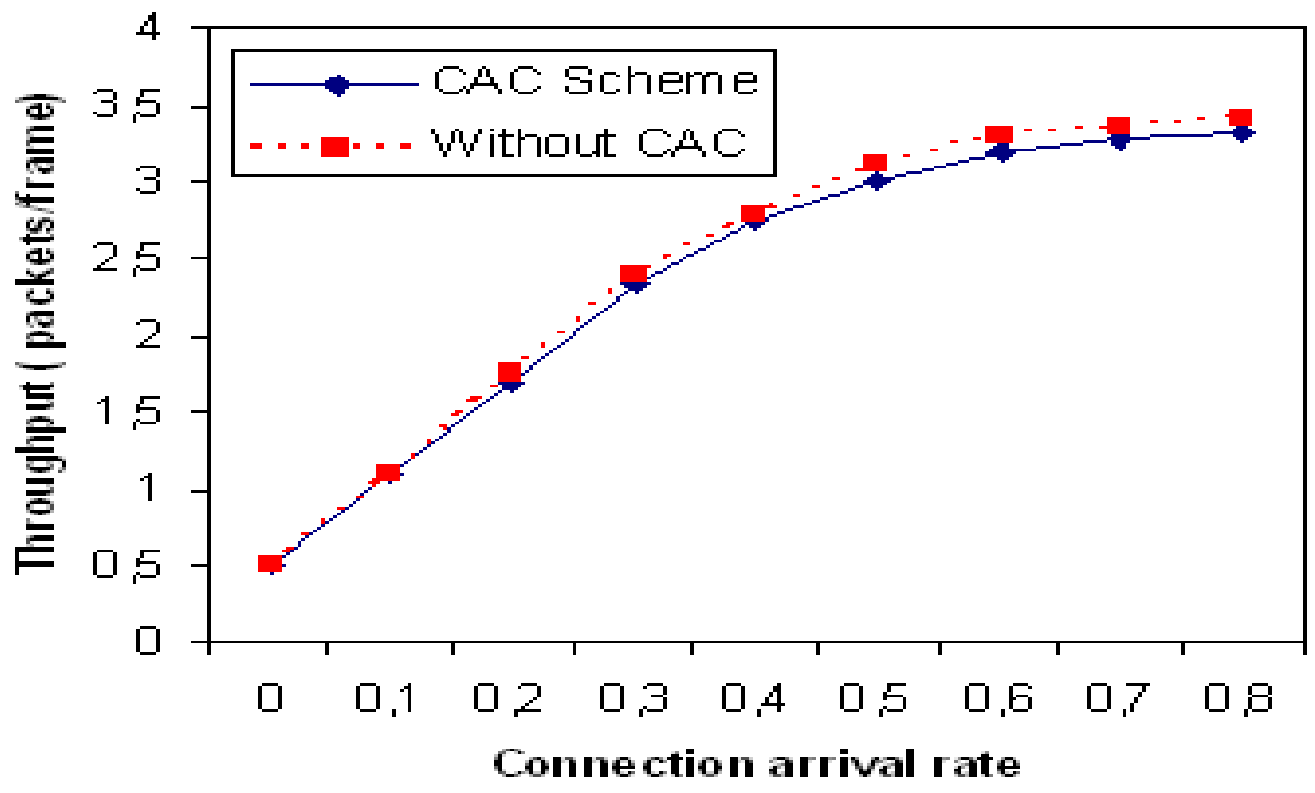

Figure 7: Queueing throughput under different connection arrival rates.

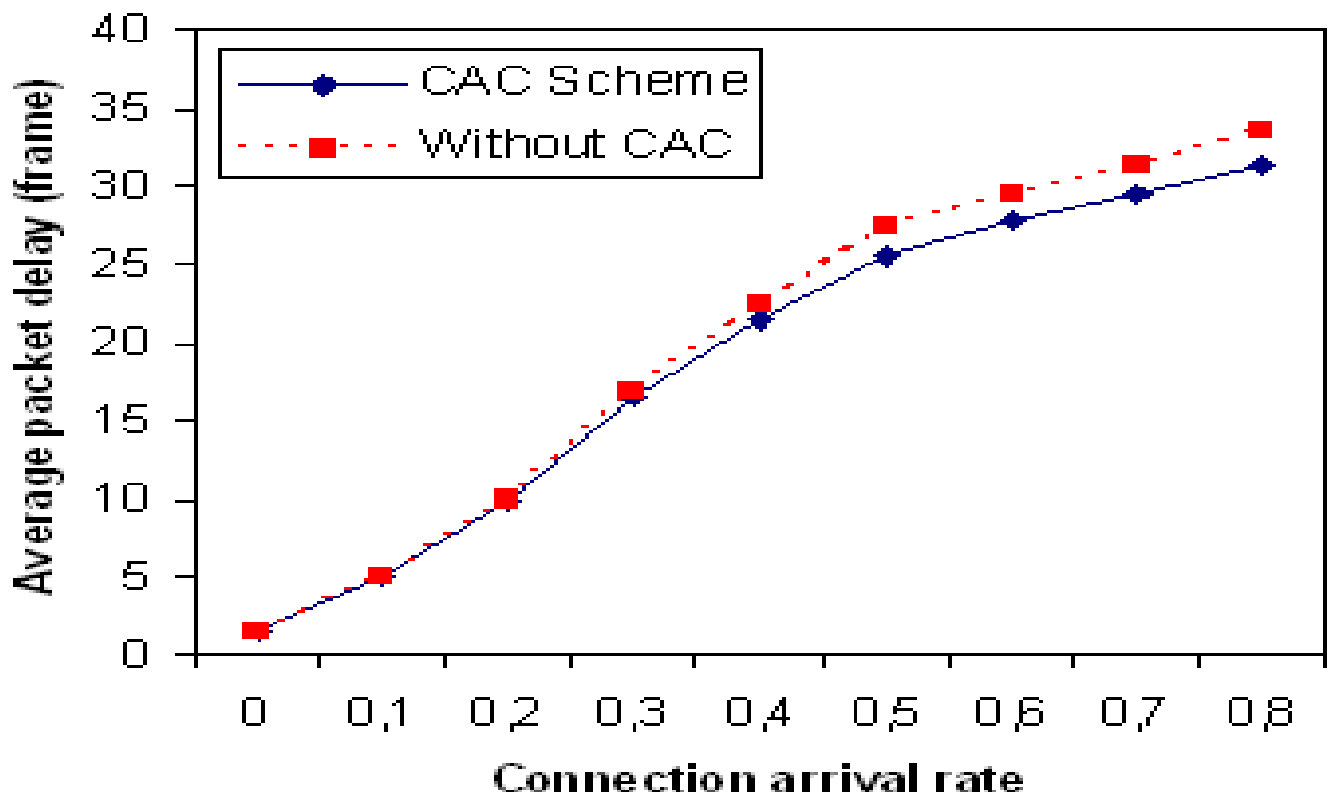

Figure 8: Average packet delay under different connection arrival rates.

Variations in packet dropping probability and average packet delay with channel quality are shown in Figures 9 and 10, respectively. As expected, the packet-level performances become better when channel quality becomes better. Also, we observe that the connection-level performances for the CAC scheme and those without CAC scheme are not impacted by the channel quality when this later becomes better (the connection blocking probability remains constant when the channel quality varies) (Figure. 11).

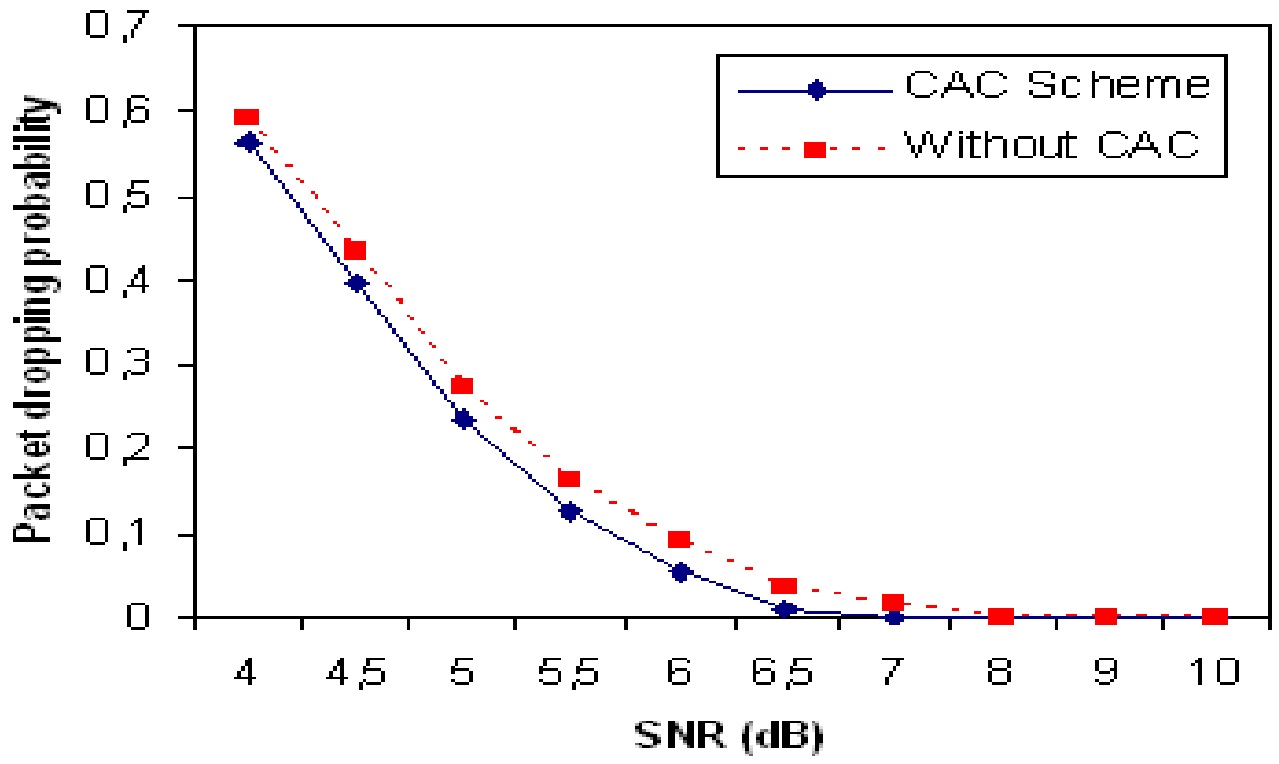

Figure 9: Packet dropping probability under different channel qualities.

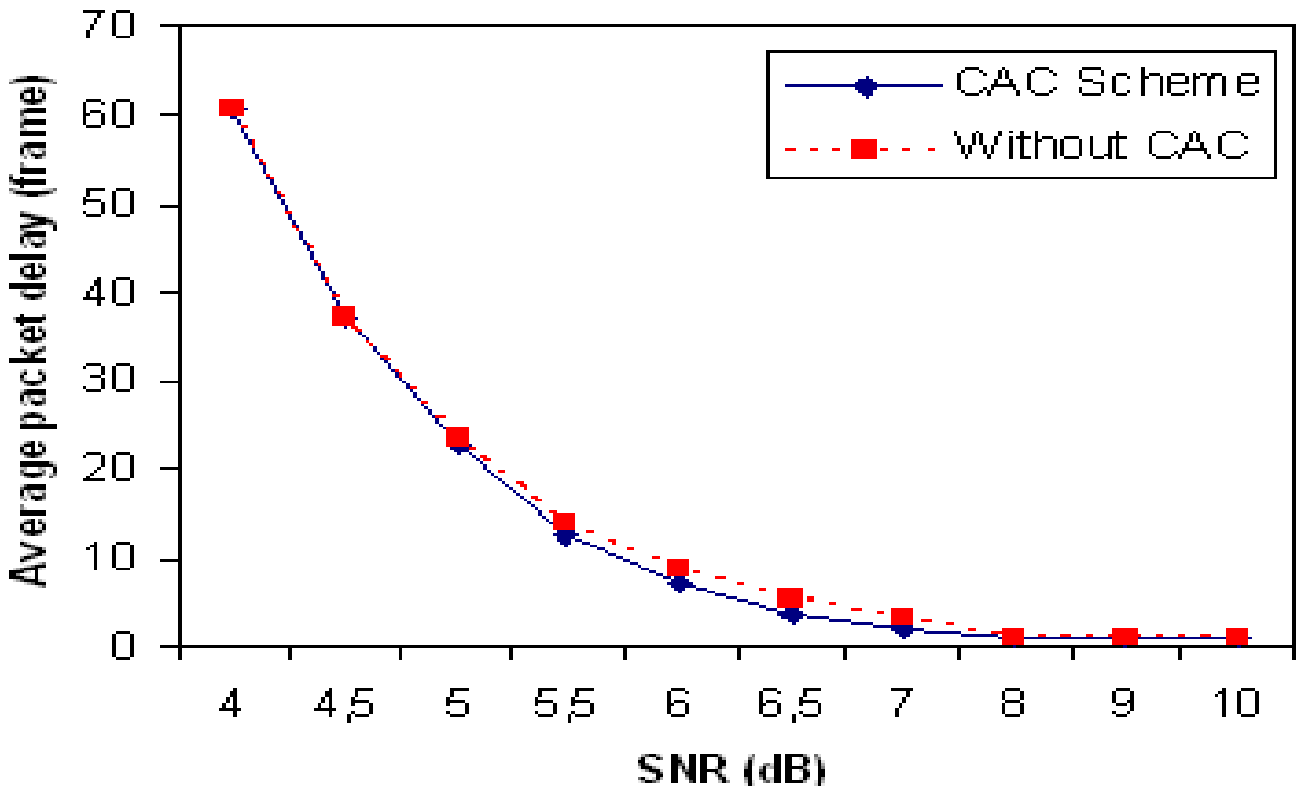

Figure 10: Average packet delay under different channel qualities

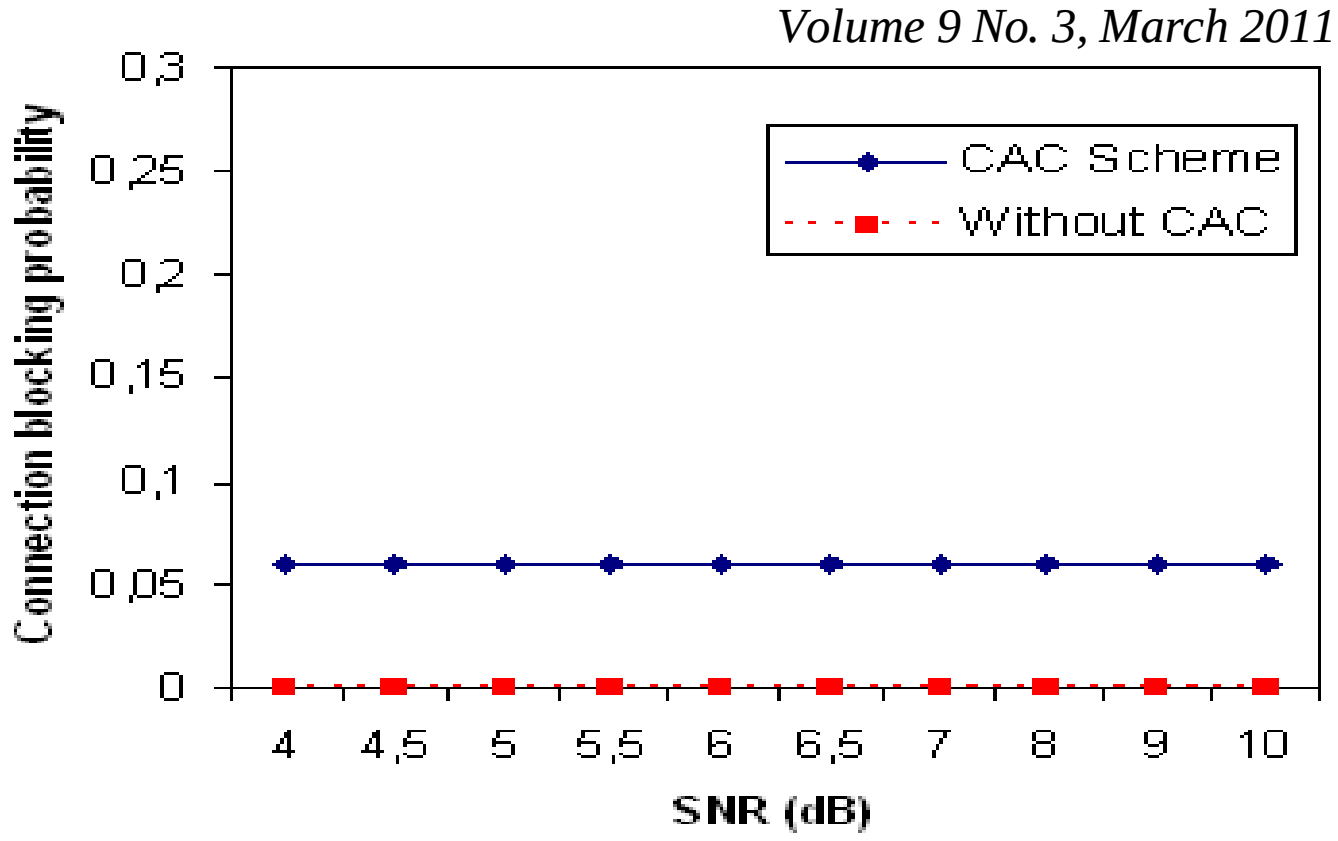

Figure 11: Connection blocking probability under different channel qualities.

## VI. Conclusion

In this paper, we have addressed the problem of queueing theoretic performance modeling and analysis of OFDMA transmission under admission control. We have considered a WiMAX system model in which a base station serves multiple subscriber stations and each of the subscriber stations is allocated with a certain number of subchannels by the base station. There are multiple ongoing connections at each subscriber station.

We have presented a connection admission control scheme for a multi-channel and multi-user OFDMA network, in which the concept of guard channel is used to limit the number of admitted connections to a certain threshold

The connection-level and packet-level performances of the CAC scheme have been studied based on the queueing model. The connection arrival is modeled by a Poisson process and the packet arrival for a connection by a two-state MMPP process. We have determined analytically and numerically different performance parameters, such as connection blocking probability, average number of ongoing connections, average queue length, packet dropping probability, queue throughput, and average packet delay.

Numerical results show that, the performance parameters of connection-level and packet-level are significantly impacted by the connection-level rate, the CAC scheme results in better packet-level performances compared with those without CAC scheme. The packet-level performances become better when channel quality becomes better. On the other hand, the connection-level performances for the CAC scheme and those without CAC scheme are not impacted by the channel quality.

All the results showed in this paper remain in correlation with those presented in [10] even if we change here the arrival packet Poisson process by an MMPP process, which is more realistic.

## References

[1] B. Baynat, S. Doirieux, G. Nogueira, M. Maqbool, and M. Coupechoux, “An efficient analytical model for wimax networks with multiple traffic profiles,” in Proc. of ACM/IET/ICST IWPAWN, September 2008.

[2] B. Baynat, G. Nogueira, M. Maqbool, and M. Coupechoux, “An efficient analytical model for the dimensioning of wimax networks,” in Proc. of 8th IFIP-TC6 Networking Conference, May 2009.

[3] L. B. Le, E. Hossain, and A. S. Alfa, “Queuing analysis for radio link level scheduling in a multi-rate TDMA wireless network,” in Proc. IEEE GLOBECOM’04, vol. 6, pp. 4061–4065, November–December 2004.

[4] Y. Fang and Y. Zhang, “Call admission control schemes and performance analysis in wireless mobile networks,” IEEE Transactions on Vehicular Technology, vol. 51, no. 2, March 2002, pp. 371–382.

[5] D. Hong and S. S. Rappaport, “Traffic model and performance analysis for cellular mobile radio telephone systems with prioritized and nonprioritized handoff procedures,” IEEE Transactions on Vehicular Technology, pp. 77–92, August 1986.

[6] H. Lee and D.-H. Cho, “VoIP capacity analysis in cognitive radio system,” IEEE Commun. Lett., vol. 13, no. 6, pp. 393–395, Jun. 2009.

[7] H. Lee, Hyu-Dae Kim and Dong-Ho Cho, "Smart Resource Allocation Algorithm Considering Voice Activity for VoIP Services in Mobile-WiMAX System," IEEE Transactions on Wireless Communications, vol 8, issue 9, pp. 4688-4697, Sep. 2009.

[8] H. Lee and Dong-Ho Cho, "Capacity Improvement and Analysis for VoIP Service in Cognitive Radio System," IEEE Transactions on Vehicular Technology, vol 59, issue 4, pp. 1646-1651, May 2010.

[9] R. Nelson, “Probability, stochastic process, and queueing theory”, Springer-Verlag, third printing, 2000.

[10] D. Niyato and E. Hossain, "Connection admission control in OFDMA-based WiMAX networks: Performance modeling and analysis," invited chapter in WiMax/MobileFi: Advanced Research and Technology, (Ed. Y. Xiao), Auerbach Publications, CRC Press, December 2007.

[11] D. Niyato and E. Hossain, “Connection admission control algorithms for OFDMA wireless networks,” in Proc. IEEE GLOBECOM’05, St. Louis, MO, USA, 28 November–2 December 2005.

[12] D. Pareek, “WiMax: Taking Wireless to the MAX,” Auerbach Publishers Inc. June, 2006.

[13] R. Ramjee, R. Nagarajan, and D. Towsley, “On optimal call admission control in cellular networks,” in Proc. IEEE INFOCOM’96, vol. 1, San Francisco, CA, March 1996, pp. 43–50.

[14] IEEE Std 802.16e-2005 and IEEE Std 802.16-2004/Cor 1-2005, IEEE Standard for Local and metropolitan area networks-Part 16: Air Interface for Fixed and Mobile Broadband Wireless Access Systems, Dec. 7, 2005.

[15] J.-W. So, “Performance analysis of VoIP services in the IEEE 802.16e OFDMA system with inband signaling,” IEEE Trans. Veh. Technol., vol.57, no. 3, pp. 1876–1886, May 2008.